\documentclass[sigconf, anonymous=False, review=False, natbib=false]{acmart}

\RequirePackage[abbreviate=true, dateabbrev=true, natbib=true, isbn=false, doi=false, urldate=comp, url=false, maxbibnames=2,  maxcitenames=1, backref=false, backend=biber, style=ACM-Reference-Format, language=american]{biblatex}

\usepackage{booktabs} 
\usepackage[hypcap=false]{subcaption}
\usepackage{multirow}
\usepackage{siunitx}
\usepackage[super]{nth}
\usepackage{amsmath}
\usepackage{algorithm}
\usepackage{cleveref}
\usepackage[normalem]{ulem}
\usepackage{array}
\usepackage{algpseudocode}

\usepackage{mleftright}
\usepackage{bbm}
\usepackage{wrapfig}
\usepackage{float}
\usepackage{xfrac}
\usepackage[flushleft]{threeparttable}
\setcopyright{rightsretained}
\copyrightyear{2018}
\acmYear{2018}
\setcopyright{acmlicensed}
\acmPrice{15.00}
\acmDOI{10.1145/1122445.1122456}
\acmISBN{978-1-4503-9999-9/18/06}

\renewcommand\footnotetextcopyrightpermission[1]{} 

\begin{document}
\newcolumntype{P}[1]{>{\centering\arraybackslash}p{#1}}
\title[]{An Induced Multi-Relational Framework for Answer Selection in Community Question Answer Platforms}


\author{Kanika Narang, Chaoqi Yang, Adit Krishnan, Junting Wang, Hari Sundaram, Carolyn Sutter}
\affiliation{%
\institution{University of Illinois, Urbana-Champaign}
}
\affiliation{%
  \institution{\{knarang2,\hspace{0.2em}, chaoqiy2, aditk2, junting3, \hspace{0.2em}hs1, carolyns\}@illinois.edu}
}

\renewcommand{\shortauthors}{Narang et al.}

\begin{abstract}

This paper addresses the question of identifying the best candidate answer to a question on Community Question Answer (CQA) forums. The problem is important because Individuals often visit CQA forums to seek answers to nuanced questions.  We develop a novel induced relational graph convolutional network (IR-GCN) framework to address the question.  We make three contributions. First, we introduce a modular framework that separates the construction of the graph with the label selection mechanism. We use equivalence relations to induce a graph comprising cliques and identify two label assignment mechanisms---label contrast, label sharing. Then, we show how to encode these assignment mechanisms in GCNs. Second, we show that encoding contrast creates discriminative magnification---enhancing the separation between nodes in the embedding space. Third, we show a surprising result---boosting techniques improve learning over familiar stacking, fusion, or aggregation approaches for neural architectures. We show strong results over the state-of-the-art neural baselines in extensive experiments on 50 StackExchange communities.

\end{abstract}

%
%
%


\maketitle

\section{Introduction}

Individuals often visit Community Question Answer (CQA) forums, like StackExchange, to seek answers to nuanced questions; typically, these answers are not readily available on web-search engines.



A fundamental challenge for a CQA search engine in response to an individual's question (i.e., information need) is to rank and identify similar past questions and relevant answers to those questions. On some CQA sites like StackExchange, individuals who post questions may label an answer as `accepted,' but other questions with answers (about $47$\% in our analysis) have none labeled as `accepted.' On other CQA sites like Reddit, there is no mechanism for a person to label an answer as `accepted.' As a first step to address the individual's information needs, in this paper, we focus on the problem of identifying accepted answers on StackExchange.

One approach to identify relevant answers is to identify salient features for each question-answer tuple $(q,a)$ and treat it as a supervised classification problem~\citep{BurelMA16,  JendersKN16, TianZL13, TianL16}. Deep Text Models further develop this approach ~ \cite{ZhangLSW17, WuWS18, WangN15, SukhbaatarSWF15}. These models learn the optimal text representation of $(q,a)$ tuple to select the most relevant answer. While the deep text models are sophisticated, text based models are computationally expensive to train. Furthermore, there are limitations to examining $(q,a)$ tuples in isolation: an answer is "relevant" \emph{in relationship} to other answers to the same question; second, it ignores the fact that same user may answer multiple questions in the forum.
An alternative approach then is to examine the graph structure resulting from users answering multiple questions in addition to the answer features. Graph Convolutional Networks (GCNs) is a popular technique to incorporate graph structure, and are used in tasks including node classification \cite{gcn} and link prediction~\cite{relationalGCN}. Extensions to the basic GCN model include signed networks~\cite{signedgcn}, inductive settings~\cite{graphsage} and multiple relations~\cite{DualGCN, relationalGCN}.

While GCNs are a plausible approach, we need to overcome a fundamental implicit assumption in prior work before we can apply it towards our problem. Prior work in GCNs adopt label sharing amongst nodes; label sharing implicitly assumes similarity between two nodes connected by an edge. In the Answer Selection problem, however, answers to the same question connected by an edge may not share acceptance label. In particular, we may label an answer as `accepted' based on how it differs with other answers to the same question. Signed GCNs~\cite{signedgcn} can not capture this contrast despite their ability to incorporate signed edges. Graph attention networks ~\cite{graphattention} also could not learn negative attention weight over neighbors as weights are the output of a softmax operation.

We develop a novel induced relational framework to address our problem. The key idea is to use diverse strategies---label depends only on the answer (reflexive), label is determined in contrast with the other answers to the question (contrastive), and label sharing among answers across questions if it contrasts with other answers similarly(similarity by contrast)---to identify the accepted answer.
Each strategy \textit{induces} a graph between $(q,a)$ tuples and then uses a particular label selection mechanism to identify the accepted answer. Our strategies generalize to a broader principle: pick an equivalence relation to induce a graph comprising cliques, and then pick a label selection mechanism (label sharing or label contrast) within each clique. We show how to develop GCN architecture to operationalize the specific label selection mechanism (label sharing or label contrast). Then, we aggregate results across strategies through a boosting framework to identify the label for each $(q,a)$ tuple. Our Contributions are as follows:
\begin{description}
  \item[Modular, Induced Relational Framework:] We introduce a modular framework that separates the construction of the graph with the label selection mechanism. In contrast,  prior work in answer selection (e.g.,~\citep{BurelMA16,  JendersKN16, TianZL13, TianL16}.) looked at individual tuples, and work on GCNs (e.g.,~\citep{gcn, DualGCN}) work with the given graph (i.e., no induced graphs) and with similarity as a mechanism for label propagation. We use equivalence relations to induce a graph comprising cliques and identify two label assignment mechanisms---label contrast, label sharing. Then, we show how to encode these assignment mechanisms in GCNs. In particular, we show that the use of equivalence relations allows us to perform \textit{exact} convolution in GCNs. We call our framework Induced Relational GCN (IR-GCN). Our framework allows for parallelization and applies to other problems that need application semantics to induce graphs independent of any existing graphs\cite{InducedGraph}.
  \item[Discriminative Semantics:] We show how to encode the notion of label contrast between a vertex and a group of vertices in GCNs. Label contrast is critical to the problem of accepted answer selection. Related work in GCNs (e.g.,~\citep{gcn, DualGCN}) emphasizes node similarity, including the work on signed graphs ~ \cite{signedgcn}. In~\citep{signedgcn}, contrast is a property of an edge, not a group and is not expressive enough for our problem. We show that our encoding of contrast creates \textit{discriminative magnification}---the separation between nodes in the embedding space is most meaningful at smaller clique sizes; the effect decreases with clique size.
  \item[Boosted Architecture:]  We show through extensive empirical results that using common boosting techniques improves learning in our convolutional model. This improvement is a surprising result since much of the work on neural architectures develops stacking, fusion, or aggregator architectures.
\end{description}

We conducted extensive experiments using our IR-GCN framework with excellent experimental results on popular CQA forum---StackExchange. For our analysis, we collect data from 50 communities---the ten largest communities from each of the five StackExchange\footnote{https://stackexchange.com/sites} categories. We achieved an improvement of over 4\% accuracy and 2.5\% in MRR on average over state-of-the-art baselines. We also show that our model is more robust to label sparsity compared to alternate GCN based multi-relational approaches.

We organize the rest of this paper as follows. In \cref{sec:problem}, we formulate our problem statement and then discuss induced relations for Answer Selection problem in \cref{sec:Induced Relational Views}. We then detail the operationalization of these induced relations in Graph Convolution framework in \cref{sec:gcn} and introduce our gradient boosting based aggregator approach in \cref{sec:aggregation}. \Cref{sec:experiments} describes experiments. We discuss related work in \cref{sec:related} and then conclude in \cref{sec:conclusion}.

\section{Problem Formulation}
\label{sec:problem}

In Community Question Answer (CQA) forums, an individual asking a question seeks to identify the most relevant candidate answer to his question. On Stack-Exchange CQA forums, users annotate their preferred answer as ``accepted.''


Let $\mathcal{Q}$ denote the set of questions in the community and for each $q \in \mathcal{Q}$, we denote $\mathcal{A}_q$ to be the associated set of answers. Each question $q \in \mathcal{Q}$, and each answer $a \in \mathcal{A}_q$ has an author $u_q, u_a \in \mathcal{U}$ respectively. Without loss of generality, assume that we can extract features for each question $q$, each answer $a \in \mathcal{A}_q$, user $u_q, u_a \in \mathcal{U}$.

Our unit of analysis is a question-answer tuple $(q,a), q \in \mathcal{Q}, a \in \mathcal{A}_q$, and we associate each $(q,a)$ tuple with a label $y_{q,a} \in \{-1,+1\}$, where `+1' implies acceptance and `-1' implies rejection.

\begin{quote}
    The goal of this paper is to develop a framework to identify the accepted answer to a question posted on a CQA forum.
\end{quote}

\section{Induced Relational Views}
\label{sec:Induced Relational Views}
In this section, we discuss the idea of induced relational views, central to our induced relational GCN framework developed in~\Cref{sec:gcn}.
First, in~\Cref{sub:Induced Views}, we introduce potential strategies for selecting the accepted answer given a question. We show how each strategy induces a graph $G$ on the question-answer $(q,a)$ tuples. Next, in~\Cref{sub:Generalized Views}, we show how each of these example strategies is an instance of an equivalence relation; our framework generalizes to incorporate any such relation.


\begin{figure}[h]
    \centering
    \begin{subfigure}{0.15\textwidth}
        \centering
        \includegraphics[scale=0.3]{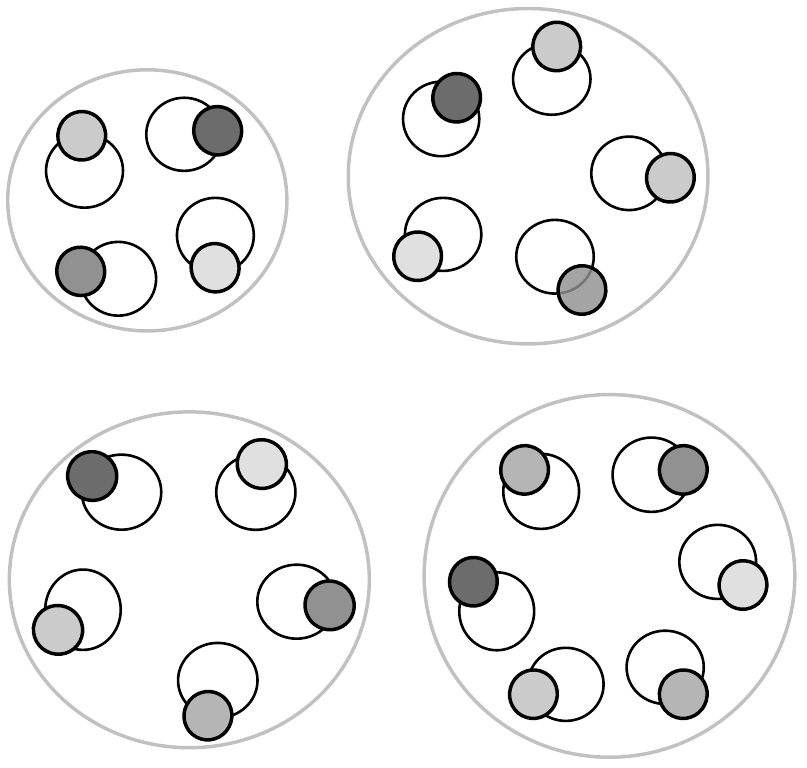}
            \caption{Reflexive}
            \label{fig:reflexive}
    \end{subfigure}%
    \begin{subfigure}{0.17\textwidth}
        \centering
        \includegraphics[scale=0.3]{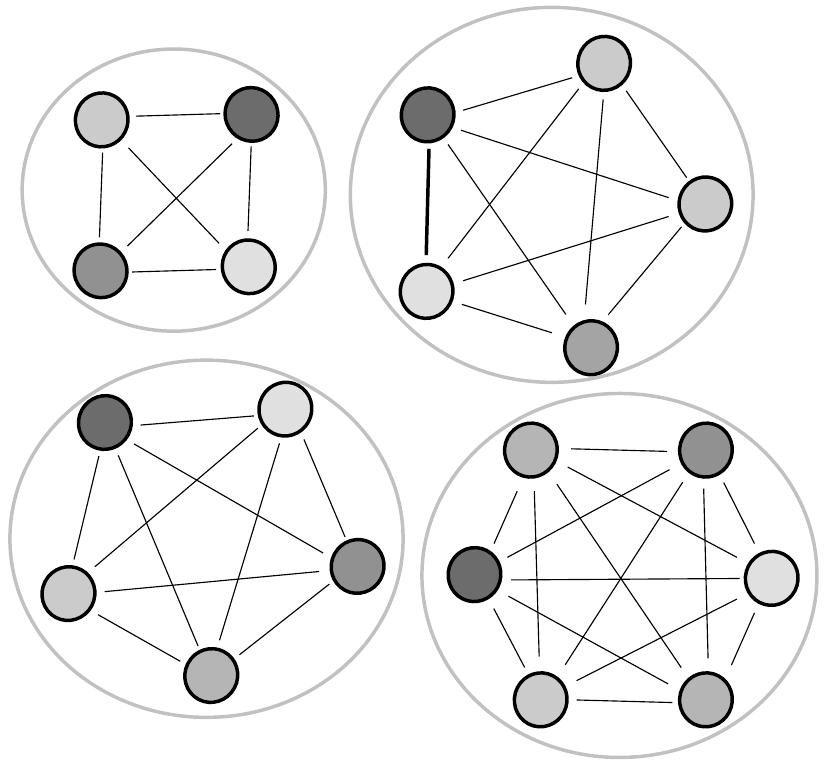}
        \caption{Contrastive}
        \label{fig:contrastive}
    \end{subfigure}%
    \begin{subfigure}{0.15\textwidth}
        \centering
            \includegraphics[scale=0.3]{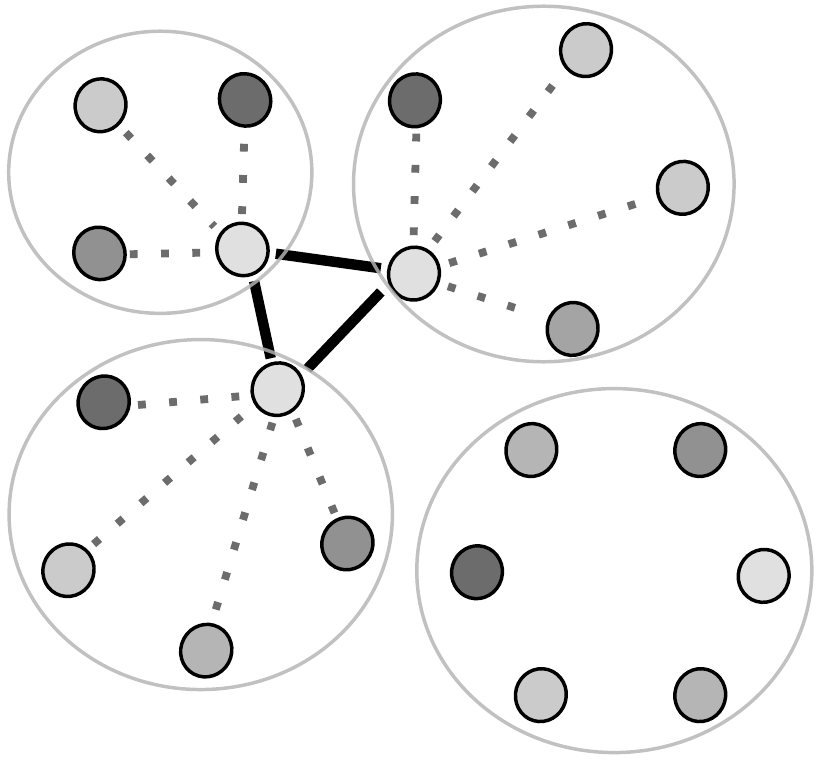}
      \caption{Similar Contrast}
      \label{fig:similar}
        \end{subfigure}%
    \caption{\small \label{fig:relation} Reflexive(~\cref{fig:reflexive}), Contrastive (~\cref{fig:contrastive}) and Similar Contrast (~\cref{fig:similar}) relations among $(q,a)$ tuples. Reflexive assumes no dependence on other answers for prediction. Contrastive compares between all answers to a question;  Similar Contrast connects answers across questions if they contrasts with other answers similarly. Solid lines show the similarity relation while dotted lines signify the contrast. The contrast is only significant in three questions.}
    \vspace{-0.15in}
\end{figure}

\subsection{Constructing Induced Views}
\label{sub:Induced Views}

In this section, we discuss in detail four example strategies that can be used by the individual posting the question to label an answer as `accepted.'
Each of the $S_i \in \mathbf{S}$ strategies \textit{induces} a graph $G_i = (V, E_i)$ (also referred to as a relational view).
In each graph $G_i$, a vertex $v \in V$ corresponds to a tuple $(q,a)$ and an edge $e \in E_i, E_i \subseteq V \times V$ connects two tuples that are matched under that strategy.
Note that each $G_i$ has the same vertex set $V$, and the edge sets $E_i$ are strategy dependent. Each strategy employs one of the three different relation types---reflexive, contrastive, and similar---to connect the tuples. We use one reflexive strategy, one contrastive, and two similar strategies.
~\Cref{fig:relation} summarizes the three relations. Below, we organize the discussion by relation type.


\subsubsection{Reflexive}
\label{sub:Reflexive}

A natural strategy is to examine each $(q,a)$ tuple in isolation and then assign a label $y_{q,a} \in \{-1,+1 \}$ corresponding to `not accepted' or `accepted.' In this case, $y_{q,a}$ depends on only the features of $(q,a)$. This is a \textbf{Reflexive} relation, and the corresponding graph $G_r = (V,E_r)$ has a specific structure. In particular, in this graph $G_r$, we have only self-loops, and all edges $e \in E_r$ are of the type $(v,v)$. That is, for each vertex $v \in V$, there are no edges $(v,u)$ to any other vertices $u\neq v \in V$. Much of the prior work on feature driven answer selection~\cite{BurelMA16,  JendersKN16, TianZL13, TianL16} adopts this view.

\subsubsection{Contrastive}
\label{sub:Contrastive}

A second strategy is to examine answers \textit{in relation} to other answers to the same question and label one such answer as `accepted.' Thus the second strategy \textit{contrasts} $(q,a)$, with other tuples in  $(q,a'), q \in \mathcal{Q}; a, a' \in \mathcal{A}_q; a'\neq a$. This is a \textbf{Contrastive} relation and the corresponding graph $G_c = (V,E_c)$ has a specific structure. Specifically, we define an edge $e \in E_c$ for all $(q,a)$ tuples for the same question $q \in \mathcal{Q}$. That is, if  $v = (q_1, a_1), u=(q_2, a_2)$, $e=(u, v) \in E_c \iff q_1=q_2$. Intuitively, the contrastive relation induces cliques connecting all answers to the same question. Introducing contrasts between vertices sharpens differences between features, an effect (described in more detail in~\Cref{subsec:contrast}) we term \emph{Discriminative Feature Magnification}. Notice that the contrastive relation is distinct from graphs with signed edges (e.g.,~\cite{signedgcn}). In our framework, the contrast is a \textit{neighborhood} property of a vertex, whereas in~\cite{signedgcn}, the negative sign is a property of an \textit{edge}.

\subsubsection{Similar Contrasts}
\label{sub:Similar}
A third strategy is to identify \textit{similar} $(q,a)$ tuples \textit{across} questions. Prior work~\cite{Wu2016} indicates that individuals on StackExchange use diverse strategies to contribute answers. Experts (with high reputation) tend to answer harder questions, while new members (with low reputation)  looking to acquire reputation tend to be the first to answer a question.

How might similarity by contrast work? Consider two individuals Alice and Bob with \textit{similar} reputations (either high or low) on StackExchange, who contribute answers $a_A$ and $a_B$ to questions $q_1$ and $q_2$ respectively. If Alice and Bob have high reputation difference with other individuals who answer questions $q_1$ and $q_2$ respectively, then it is likely that $(q_1, a_A)$ and $(q_2, a_B)$ will share the same label (if they are both experts, their answers might be accepted, if they are both novices, then this is less likely). However, if Alice has a high reputation difference with other peers who answer $q_1$, \textit{but Bob does not have that difference} with peers who answer $q_2$, then it is less likely that the tuples $(q_1, a_A)$ and $(q_2, a_B)$ will share the label, even though the reputations of Alice and Bob are similar.

Thus the key idea of the \textbf{Similar Contrasts} relation is that link tuples that are  \textit{similar in how they differ} with other tuples. We construct the graph $G_s = (V, E_s)$ in the following manner. An edge $e = (v,u)$ between tuples $v$ and $u$ exists if the similarity $s(v,u)$ between tuples $v,u$ exceeds a threshold $\delta$. We define the similarity function $s(\cdot , \cdot)$ to encode similarity by contrast. That is, $e=(v,u) \in E_s \iff s(v,u) \geq \delta$.


Motivated by~\cite{Wu2016}, we consider two different views that correspond to the similar contrast relation. The \textbf{TrueSkill Similarity} view connects all answers authored by a user where her skill (computed via Bayesian TrueSkill~\cite{TrueSkill06})) differs from competitors by margin $\delta$. We capture both cases when the user is less or more skilled than her competitors. In the \textbf{Arrival Similarity} view, we connect answers across questions based on the similarity in the relative time of their arrival (posting timestamp).
Notice that two Similar Contrast views have different edge ($E$) sets since the corresponding similarity functions are different. Notice also, that the two similarity function definitions are transitive.
 \footnote{One trivial way of establishing similarity is co-authorship i.e. connect all $(q,a)$ tuples of a user (probably on the same topic) across different questions.
Note that the accepted answer is labeled in relation to the other answers. As the competing answers are different in each question, we can not trivially assume acceptance label similarity for all co-authored answers. In our experiments, co-authorship introduced a lot of noisy links in the graph leading to worse performance.}

\subsection{Generalized Views}
\label{sub:Generalized Views}
Now we present the general case of the induced view. First, notice that each of the three relation types that we consider---reflexive, contrastive, and similar---result in a graph $G_i = (V, E_i)$ comprising a set of cliques. This is not surprising, since all three relations presented here, are equivalence relations. Second, observe the semantics of how we select the tuple with the accepted answer. Within the three relations, we used two semantically different ways to assign the `accepted' answer label to a tuple. One way is to share the labels amongst all the vertices in the \textit{same clique} (used in the reflexive and the similar relations). Second is to \textit{assign label based on contrasts with other vertices} in the same clique. We can now state the organizing principle of our approach as follows.
\begin{quote}
  A generalized \textit{modular} framework: pick a meaningful equivalence relation on the $(q,a)$ tuples to induce graph comprising cliques and then apply specific label semantics within each clique.
\end{quote}

Equivalence relation results in a graph with a set of disconnected cliques. Then, within a clique, one could use application-specific semantics, different from two discussed in this paper, to label tuples as `accepted.'
Cliques have some advantages: they have well-defined graph spectra~\cite[p. 6]{Chung1997}; cliques allows for \textit{exact} graph convolution; parallelize the training as the convolution of a clique is independent of other cliques.

Thus, each strategy induces a graph $G_i=(V,E_i)$ using one of the three equivalence relations---reflexive, contrastive and similar---and then applies one of the two semantics (`share the same label'; `determine label based on contrast').

\section{Induced Relational GCN}
\label{sec:gcn}
Now, we will encode the two label assignment mechanisms within a clique
via a graph convolution. First, we briefly review Graph Convolution Networks (GCN) and identify some key concepts. Then, given the views $G_i$ for the four strategies, we show how to introduce label contrasts in~\Cref{subsec:contrast} followed by label sharing in~\Cref{subsec:similar}.


\subsection{Graph Convolution}
\label{subsec:graph}
Graph Convolution models adapt the convolution operations on regular grids (like images) to irregular graph-structured data $G = (V,E)$, learning low-dimensional vertex representations. If for example, we associate a scalar with each vertex $v \in V$, where $|V| = N$, then we can describe the convolution operation on a graph by the product of signal $x \in \mathbb{R}^N$ (feature vectors) with a learned filter $g_\theta$ in the fourier domain. Thus,
\begin{equation}
  g_\theta \ast x =  U \, g_\theta \, U^T x,
  \label{eq:basic_gcn}
\end{equation}
where, $\Lambda$ and $U$ are the eigenvalues and eigenvector of the normalized graph Laplacian, $L = I_N - D^{-\sfrac{1}{2}}AD^{\sfrac{1}{2}}$, and where $L = U \Lambda U^T$. $A$ denotes the adjacency matrix of a graph $G$ (associated with a view) with $N$ vertices. ~\Cref{eq:basic_gcn} implies a filter $g_\theta$ with $N$ free parameters, and requires expensive eigenvector decomposition of the adjacency matrix $A$. ~\citet{deferrard} proposed to approximate $g_\theta$, which in general is a function of $\Lambda$, by a sum of Chebyshev polynomials $T_k(x)$ up to the $k$-th order. Then,

\begin{equation}
  g_\theta \ast x \approx U \, \sum_{k=0}^K \theta_k T_k(\tilde{\Lambda}) \, U^T x \approx \, \sum_{k=0}^K \theta_k T_k(\tilde{L}) \, x,
  \label{eq:approx_gcn}
\end{equation}
where, $\tilde{\Lambda} = 2 \Lambda/ \lambda_{\max}- I_N$ are the scaled eigenvalues and $\tilde{L} = 2L/\lambda_{max} - I_N$ is the corresponding scaled Laplacian. Since $\tilde{L} = U \tilde{\Lambda} U^T$, the two equations are approximately equal.


The key result from~\citet{deferrard} is that~\Cref{eq:approx_gcn} implies $k$-hop localization---the convolution result depends only on the $k$-hop neighborhood. In other words,~\Cref{eq:approx_gcn}  is a $k$-hop approximation.

However, since we use equivalence relations in our framework that result in cliques, we can do an \textit{exact} convolution operation since vertices in a clique only have one-hop (i.e., $k=1$) neighbors (see lemma 5.2, \cite{Hammond2011}). The resulting convolution is linear in $L$ and now has only two filter parameters, $\theta_{0}$ and $\theta_{1}$ shared over the whole graph.
\begin{equation}
g_{\theta} * x = \theta_{0}x + \theta_{1}\left(L-I_{N} \right)x 
\label{eq:restrictk}
\end{equation}

We emphasize the distinction with~\citet{gcn} who approximate the~\citet{deferrard} observation by restricting $k=1$. They do so since they work on arbitrary graphs; since our relations result in views with cliques, we do not make any approximation by using $k=1$.

\subsection{Contrastive Graph Convolution}
\label{subsec:contrast}

Now, we show how to perform graph convolution to encode the mechanism of contrast, where label assignments for a tuple depend on the contrast with its neighborhood.

To establish contrast, we need to compute the \emph{difference} between the vertex's own features to its neighborhood in the clique. Thus we transform~\Cref{eq:restrictk} by setting $\theta = \theta_{0}$ = $\theta_{1}$, which essentially restricts the filters learned by the GCN. This transformation leads to the following convolution operation:
\begin{align}
g_{\theta} * x & =  \theta \left( I_N + L- I_{N} \right) x \\
g_{\theta} * x & =  \theta \left( I_N - D^{-\sfrac{1}{2}} A D^{-\sfrac{1}{2}}\right) x \label{eq:contrastdetail}
\end{align}

Notice that~\Cref{eq:contrastdetail} says that for example, for any vertex $u$ with a scalar feature value $x_u$, for a given clique with $n \geq 2$ vertices, the convolution operation computes a new value $\hat{x}_u$ for vertex $u$ as follows:
\begin{equation}
  \hat{x}_u = \theta \left ( x_u - \frac{1}{n-1} \sum_{v \in \mathcal{N}_u} x_v \right ).
\end{equation}
where $\mathcal{N}_u$ is the neighborhood of vertex $u$. Notice that since our equivalence relations construct cliques, for all vertices $u$ that belong to a clique of size $n$, $|\mathcal{N}_u| = n-1$.

When we apply the convolution operation in~\Cref{eq:contrastdetail} at each layer of GCN, output for the $k$-th layer is:

\begin{equation}
  \label{eq:contrast}
  \mathbf{Z}_c^{k} = \sigma \left( \left (I_N - D^{-\sfrac{1}{2}}A_cD^{\sfrac{1}{2}} \right) \mathbf{Z}_c^{k-1} \mathbf{W}_c^{k}\right)
\end{equation}
with $A_c$ denoting the adjacency matrix in the contrastive view. $\mathbf{Z}_c^{k} \in \mathbb{R}^{N \times d}$ are the learned vertex representations for each $(q,a)$ tuple under the contrastive label assignment. $N$ is the total number of tuples and $d$ refers to the dimensionality of the embedding space. $\mathbf{Z}^{k-1}$ refers to the output of the previous $(k-1)$-{th} layer, and $\mathbf{Z}^{0} = X$ where $X$ is the input feature matrix. $\mathbf{W}_c^{k}$ are the filter $\theta$ parameters learnt by the GCN; $\sigma( \cdot)$ denotes the activation function (e.g. ReLU, $\tanh$).


To understand the effect of~\Cref{eq:contrast} on a tuple, let us restrict our attention to a vertex $u$ in a clique of size $n$. We can do this since the convolution result in one clique is unaffected by other cliques. When we do this, we obtain:
\begin{equation}
  z_c^{k}(u) = \sigma \left(\left(z_c^{k-1}(u) - \frac{1}{n-1} \sum_{v \in \mathcal{N}_u} z_c^{k-1}(v) \right) \mathbf{W}_{c}^{k}\right). \label{eq:contrastrestrict}
  \end{equation}

Now consider a pair of contrasting vertices, $u$ and $v$ in the same clique of size $n$. Let us ignore the linear transform by setting $W_{c}^{k}=\mathbf{I}$ and set $\sigma(\cdot)$ to the identity function. Then we can easily verify that:

\begin{equation}
z_c^{k}(u) - z_c^{k}(v) = \underbrace{
  \left (1 + \frac{1}{n-1} \right )
  }_{\text{magnification}}
  \times
  \underbrace{
    \left ( z_c^{k-1}(u) - z_c^{k-1}(v) \right )
    }_{\text{contrast in previous layer}}, \label{eq:disccontrastsimple}
\end{equation}
where, $z_c^{k}(u)$ denotes the output of the $k$-th convolution layer for the $u$-th vertex in the contrastive view. As a result, each convolutional layer magnifies the feature contrast between the vertices that belong to the same clique. Thus, the contrasting vertices move further apart. We term this as \emph{Discriminative Feature Magnification} and~\Cref{eq:disccontrastsimple} implies that we should see higher magnification effect for smaller cliques.

\subsection{Encoding Similarity Convolution}
\label{subsec:similar}
We next discuss how to encode the mechanism of sharing labels in a GCN. While label sharing applies to our similar contrast relation (two strategies: Arrival similarity; TrueSkill similarity, see~\Cref{sub:Induced Views}), it is also trivially applicable to the reflexive relation, where the label of the tuple only depends on itself. First, we discuss the case of similar contrasts.

\subsubsection{Encoding Similar Contrasts}
\label{sub:Encoding Similar Contrasts}

To encode label sharing for the two similar by contrast cases, we transform~\Cref{eq:restrictk} with the assumption $\theta = \theta_0 = -\theta_1$. Thus

\begin{equation}
g_{\theta} * x = \theta\left(I_{N} + D^{-\sfrac{1}{2}}AD^{-\sfrac{1}{2}}\right) x, \label{eq:similargcn}
\end{equation}

Similar to the ~\Cref{eq:contrastdetail} analysis, convolution operation in \Cref{eq:similargcn} computes a new value $\hat{x}_u$ for vertex $u$ as follows:
\begin{align}
  \hat{x}_u &= \theta \left ( x_u + \frac{1}{n-1} \sum_{v \in \mathcal{N}_u} x_v \right ).\\
  \hat{x}_u &= \theta \left ( \frac{n-2}{n-1} x_u + \frac{n}{n-1} \mu_x \right ).
\end{align}
That is, in the mechanism where we share labels in a clique, the convolution pushes the values of each vertex in the clique to the average feature value, $\mu_x = \frac{1}{n} \sum_{v \in \mathcal{N}_u \cup u} x_v$, in the clique.

When we apply the convolution operation in~\Cref{eq:similargcn} at each layer of GCN, output for the $k$-th layer:

\begin{equation}
  \label{eq:similar}
  \mathbf{Z}_s^{k} = \sigma \left( \left (I_N + D^{-\sfrac{1}{2}}A_sD^{\sfrac{1}{2}} \right) \mathbf{Z}_s^{k-1} \mathbf{W}_s^{k}\right)
\end{equation}
with $A_s$ denoting the adjacency matrix in the similar views.

We analyze the similarity GCN in a maner akin to~\Cref{eq:contrastrestrict} and
we can easily verify that:

\begin{equation}
z_s^{k}(u) - z_s^{k}(v) = \underbrace{
  \left (1 - \frac{1}{n-1} \right )
  }_{\text{reduction}}
  \times
  \underbrace{
    \left ( z_s^{k-1}(u) - z_s^{k-1}(v) \right )
    }_{\text{contrast in previous layer}}, \label{eq:diffsimilar}
\end{equation}
where, $z_s^{k}(i)$ denotes the output of the $k$-th convolution layer for the $i$-th vertex in the similar view. As a result, each convolutional layer reduces the feature contrast between the vertices that belong to the same clique. Thus, the similar vertices move closer.



The proposed label sharing encoding applies to both similar contrast strategies (TrueSkill; Arrival). We refer to the corresponding vertex representations as $\mathbf{Z}_{ts}^{k}$ (TrueSkill), $\mathbf{Z}_{as}^{k}$ (Arrival).

\subsubsection{Reflexive Convolution}
\label{subsubsec:reflex}
We encode the reflexive relation with self-loops in the graph resulting in an identity adjacency matrix. This relation is the trivial label sharing case, with an independent assignment of vertex labels. Thus, the output of the $k$-th convolutional layer for the reflexive view, $\mathbf{Z}_r^{k}$ reduces to:
\begin{equation}
  \label{eq:reflexive}
  \mathbf{Z}_r^{k} = \sigma \left( I_N \mathbf{Z}_r^{k-1} \mathbf{W}_r^{k} \right)
\end{equation}
Hence, the reflexive convolution operation is equivalent to a feedforward neural network with multiple layers and activation $\sigma( \cdot )$.

\vspace{0.1in}
\noindent
Each strategy $S_i \in \mathbf{S}$ belongs to one of the three relation types---reflexive, contrastive and similarity, where $\mathbf{R}$ denotes the set of strategies of that relation type. $\mathcal{R} = \bigcup \mathbf{R}$ denotes the set of all relation types.
$\mathbf{Z}_i^K \in \mathbb{R}^{N X d}$ represents the $d$ dimensional vertex embeddings for strategy $S_i$ at the $K$-th layer. For each strategy $S_i$, we obtain a scalar score by multiplying $\mathbf{Z}_i^K$ with transform parameters $\widetilde{W}_i \in \mathbb{R}^{d \times 1}$.
The sum of these scores gives the combined prediction score, $\mathbf{H}_{\mathbf{R}} \in \mathbb{R}^{N X 1}$, for that relation type.
\begin{equation}
    \label{eq:score}
        \mathbf{H}_{\mathbf{R}} = \sum_{S_i \in \mathbf{R}} \mathbf{Z}_i^K \widetilde{W}_i^T
\end{equation}


In this section, we proposed Graph Convolutional architectures to compute vertex representations of each $(q,a)$ tuple under the four strategies. 
In particular, we showed how to encode two different label assignment mechanisms---label sharing and determine label based on contrast---within a clique. The architecture that encodes label assignment based on contrast is a novel contribution; distinct from the formulations presented by~\citet{gcn} and its extensions~\cite{signedgcn, relationalGCN}. Prior convolutional architectures implicitly encode the label sharing mechanism (~\cref{eq:similargcn}); however, label sharing is unsuitable for contrastive relationships across vertices. Hence our architecture fills this gap in prior work.


\section{Aggregating Induced Views}
\label{sec:aggregation}
In the previous sections, we introduced four strategies to identify the accepted answer to a question. Each strategy induces a graph or relational view between $(q,a)$ tuples.
Each relational view is expected to capture semantically diverse neighborhoods of vertices. The convolution operator aggregates the neighborhood information under each view. The key question that follows is, \emph{how do we combine these diverse views in a unified learning framework?} Past work has considered multiple solutions:
\begin{itemize}
  \label{item:aggregator}
\item \textbf{Neighborhood Aggregation}: In this approach, we represent vertices by aggregating feature representations of it's neighbors across all views \cite{graphsage, relationalGCN}.
\item \textbf{Stacking}: Multiple convolution layers stacked end-to-end (each potentially handling a different view) \cite{Stacking}.
\item \textbf{Fusion}: Follows a multi-modal fusion approach~\cite{Fusion18}, where views are considered distinct data modalities.
\item \textbf{Shared Latent Structure}: Attempts to transfer knowledge across relational views (modalities) with constraints on the representations (e.g. \cite{DualGCN} aligns embeddings across views).
\end{itemize}

Ensemble methods introduced in \cite{relationalGCN} work on multi-relational edges in knowledge graphs. None of these approaches are directly suitable for our induced relationships. Our relational views utilize different label assignment semantics (label sharing within a clique vs. determine label based on contrast within a clique). In our label contrast semantics, we must achieve feature discrimination and label inversion between contrasting vertices, as opposed to label homogeneity and feature sharing in the label sharing case. Thus, aggregating relationships by pooling, concatenation, or addition of vertex representations fail to capture semantic heterogeneity of the induced views.
Further, data induced relations are uncurated and inherently noisy. Directly aggregating the learned representations via Stacking or Fusion can lead to noise propagation. We also expect views of the same relation type to be correlated.

\begin{figure}
    \centering
    \includegraphics[scale=0.4]{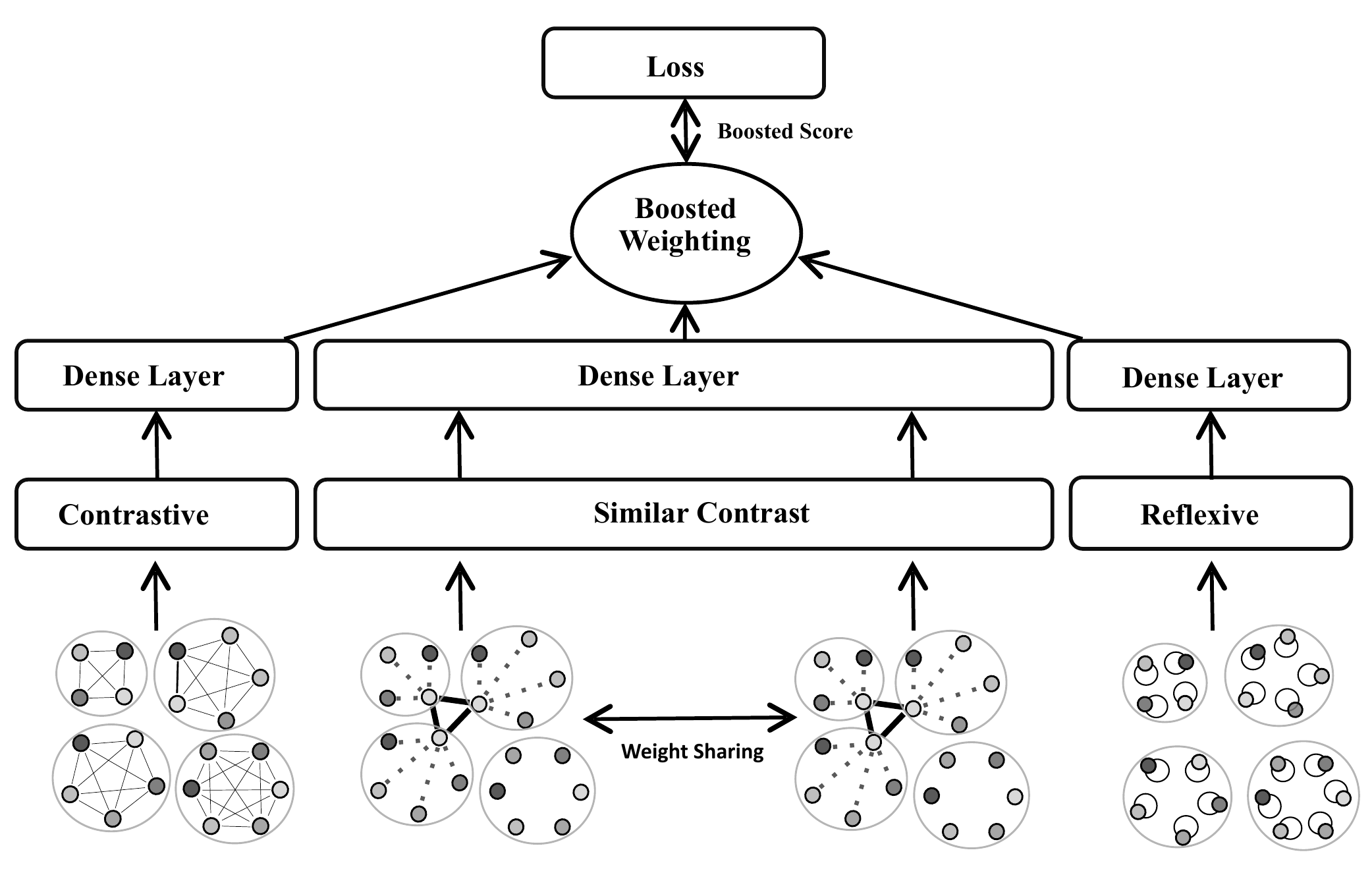}
    \caption{\small \label{fig:adaboost} Schematic diagram of our proposed IR-GCN model.} 
    \vspace{-0.1in}
\end{figure}


We thus propose the following approach to aggregate information across relation types and between views of a relation type.

\noindent
\textbf{Cross-relation Aggregation}: We expect distinct relation types to perform well on different subsets of the set of $(q,a)$ tuples. We empirically verify this with the Jaccard overlap between the set of misclassified vertices under each relational view of a relation type on our dataset. Given $\mathbf{M}_A$ and $\mathbf{M}_B$, the sets of $(q,a)$ tuples misclassified by GCNs $A$ and $B$ respectively, the jaccard overlap is,
\begin{equation*}
 \mathcal{J}_{A,B} = \frac{\mathbf{M}_A \cap \mathbf{M}_B}{\mathbf{M}_A \cup \mathbf{M}_B}
\end{equation*}
The $\mathcal{J}_{A,B}$ values are as follows for the relational pairings: (Contrastive, TrueSkill Similarity) = 0.42, (Contrastive, Reflexive) = 0.44 and (Reflexive, TrueSkill Similarity) = 0.48. Relatively low values of the overlap metric indicate uncorrelated errors across the relations.

Gradient boosting techniques are known to improve performance when individual classifiers, including neural networks \cite{ncboost}, are diverse yet accurate. A natural solution then is to apply boosting to the set of relation types and bridge the weaknesses of each learner. We employ Adaboost \cite{adaboost} to combine relation level scores, $\mathbf{H}_{\mathbf{R}}$ (~\cref{eq:score}) in a weighted manner to compute the final boosted score, $\mathbf{H}_b \in \mathbb{R}^{N \times 1}$ representing all relation types (Line 12, ~\cref{alg:inference}). $\mathbf{Y} \in \mathbb{R}^{N X 1}$ denotes the acceptance label of all tuples. Note that an entry in $(\mathbf{Y} \odot \mathbf{H_{\mathbf{R}}}) > 0 $ when the accepted label of the corresponding $(q,a)$ tuple and sign of the prediction score, $sign(\mathbf{H_{\mathbf{R}}})$, of relation type $\mathbf{R}$ match and $< 0$ otherwise. Thus, the weights $\alpha_\mathbf{R}$ adapt to the fraction of correctly classified tuples to the misclassified tuples by the relation $\mathbf{R}$ (Line 9, ~\cref{alg:inference}).
The precise score computation is described in ~\cref{alg:inference}. We use the polarity of each entry in the boosted score, $sign(\mathbf{H}_b) \in \{-1,1 \}$, to predict the class label of the corresponding $(q,a)$ tuple. The final score is also used to create a ranked list among all the candidate answers, $a \in \mathcal{A}(q)$ for each question, $q \in \mathcal{Q}$. $L_{(q,a)}$ represents the position of candidate answer $a$ in the ranked list for question $q$.

\begin{algorithm}
\caption{IR-GCN Boosted Score Computation}\label{alg:inference}
\begin{algorithmic}[1]
\Function{Forward}{$\mathbf{X}, \mathbf{Y}, \{A_i\}_{S_i \in \mathbf{S}}$}
  \State $\mathbf{H}_{b} \gets \mathbf{0} $
    \For{$\mathbf{R} \in \mathcal{R}$}
    \State $\{ \mathbf{Z}_i^K \}_{S_i \in \mathbf{R}} \gets Conv(\mathbf{X}, \{ A_i \}_{S_i \in \mathbf{R}})$
    \State \Comment{Equation  \ref{eq:contrast}, \ref{eq:similar}, \ref{eq:reflexive}}
    \State $\mathbf{H}_\mathbf{R} =\sum_{{S_i} \in \mathbf{R}} \mathbf{Z}_i^{K} \times \widetilde{\mathbf{W}}_i$ \Comment{Equation \ref{eq:score}}
    \State $ \mathbf{e}_{\mathbf{R}} \gets \exp({-\mathbf{Y} \odot \mathbf{H}_{b}})$
    \State \Comment{ $\odot \rightarrow \textit{Hadamard Product}$}
  \State $\alpha_\mathbf{R} \gets \dfrac{1}{2} \ln{\dfrac{\sum \mathbf{e}_{\mathbf{R}} \odot \mathbbm{1}\left((\mathbf{Y} \odot \mathbf{H}_{\mathbf{R}}\right) > 0)}{\sum \mathbf{e}_{\mathbf{R}} \odot \mathbbm{1}\left((\mathbf{Y} \odot \mathbf{H}_{\mathbf{R}}) < 0 \right) }}$
    \State \Comment{$\sum \rightarrow \textit{reduce-sum}$}
      \State \Comment{$\mathbbm{1}(.) \rightarrow \textit{element-wise Indicator function}$}
  \State    $\mathbf{H}_{b} \gets \mathbf{H}_{b} + \alpha_\mathbf{R} * \mathbf{H}_{\mathbf{R}}$ \Comment{Update boosted GCN}
    \EndFor
    \State \Return $\mathbf{H}_{b}$, $ \{ \mathbf{H}_{R} \}_{\mathbf{R} \in \mathcal{R}}$, $\{ \mathbf{Z}_{i}^{K} \}_{S_i \in \mathbf{S}}$
    \State \Comment{Boosted scores, Relation level scores,}
    \State \Comment{Each GCN vertex representations}
\EndFunction
\end{algorithmic}
\end{algorithm}


\noindent
\textbf{Intra-relation Aggregation}: Gradient boosting methods can effectively aggregate relation level representations, but are not optimal within a relationship type (since it cannot capture shared commonalities between different views of a relation type). For instance, we should facilitate information sharing between the TrueSkill similarity and Arrival similarity views. Thus, if an answer is authored by a user with a higher skill rating and answered significantly earlier than other answers, its probability to be accepted should be mutually enhanced by both signals. Empirically, we also found True Skill and Arrival Similarity GCNs to commit similar mistakes ($\mathcal{J}_{TS,AS}$ = 0.66). Thus, intra-relation learning (within a single relation type like Similar Contrast) can benefit from sharing the structure of their latent spaces i.e., weight parameters of GCN.




\noindent
\emph{Weight Sharing:} For multiple views representing a relation type (e.g., TrueSkill and Arrival Similarity), we train a separate GCN for each view but share the layer-wise linear-transforms $\mathbf{W}_i^{k}$ to capture similarities in the learned latent spaces.
Weight sharing is motivated by a similar idea explored to capture local and global views in \cite{DualGCN}. Although sharing the same weight parameters, each GCN can still learn distinct vertex representations as each view convolves over a different neighborhood and employ random dropout during training.
We thus propose to use an alignment loss term to minimize prediction difference between views of a single relation type\cite{reg}. The loss attempts to align the learned vertex representations at the \emph{last layer} $K$ (the loss term aligns pairs of final vertex representations, $\lvert\lvert \mathbf{Z}_i^{K} - \mathbf{Z}_{i'}^{K} \lvert\lvert \texttt{  }\forall\texttt{ } S_i, S_i' \in \mathbf{R}$). In principle, multiple GCNs augment performance of the relation type by sharing prior knowledge through multiple Adjacency matrices ($\mathbf{A}_i \texttt{  }\forall\texttt{ } S_i \in \mathbf{R}$).

\begin{algorithm}[tbh]
\caption{IR-GCN Training}\label{alg:training}
\begin{algorithmic}[1]
\Require{Input Feature Matrix $X$, Acceptance labels for each tuple, $\mathbf{Y}$, Adjacency matrix of each view $\{A_i\}_{S_i \in \mathbf{S}}$ }
\Ensure{Trained Model i.e. Weight parameters $W_{i}^{1} \ldots W_{i}^{k}, S_i \in \mathbf{S}, \forall k \in [1, K]$ and transform parameters $\widetilde{W}_i$, $S_i \in \mathbf{S}$ }
\For{$t \gets 1$ to $\textit{num-epochs}$}
    \State $\mathbf{H}_b, \{ \mathbf{H}_{R} \}_{\mathbf{R} \in \mathcal{R}}, \{ \mathbf{Z}^{K}_{i} \}_{S_i \in \mathbf{S}}$$\gets \textsc{Forward}(X, Y, \{ A_i \}_{S_i \in \mathbf{S}})$
    \State \Comment{\Cref{alg:inference}}
    \For{ $\mathbf{R} \in \mathcal{R}$}

        \State $\mathcal L_b \gets \sum \exp({-\mathbf{Y} \odot \mathbf{H}_b}) + \gamma_1 \mathcal L_1(.) + \gamma_2 \mathcal L_2(.)$
    \State \Comment{$\sum \rightarrow \textit{reduce-sum}$}
  \State \Comment{$\odot \rightarrow \textit{Hadamard Product}$}
        \State $\mathcal L_{\mathbf{R}} \gets 0$
        \For{ $S_i \in \mathbf{R}$}
        \State $\mathcal L_{i} \gets \sum \exp({-\mathbf{Y} \odot \mathbf{H}_\mathbf{R}})$
    \State $\mathcal L_{\mathbf{R}} \gets \mathcal L_{\mathbf{R}} + \mathcal L_{i} + \frac{1}{2}\sum_{S_i' \neq S_i}\lvert\lvert \mathbf{Z}_{i}^K - \mathbf{Z}_{i'}^K \lvert\lvert $
    \EndFor
    \State $\mathcal L_b \gets \mathcal L_b + \lambda(t) \mathcal L_{\mathbf{R}}$
    \State    $W_i^{k} \gets  W_i^{k} + \eta_{\textsc{adam}} \frac{\partial \mathcal L_b}{\partial W_i^{k}} $ \Comment{$\forall k \in [1, K], \forall S_i \in \mathbf{R}$}
     \State    $\widetilde{W}_i \gets  \widetilde{W}_i +  \eta_{\textsc{adam}} \frac{\partial \mathcal L_b}{\partial \widetilde{W}_i}$ \Comment{$\forall S_i \in \mathbf{S}$}
    \EndFor
\EndFor
\end{algorithmic}
\end{algorithm}

\noindent
\textbf{Training Algorithm}: Algorithm \ref{alg:training} describes the training algorithm for our IR-GCN model. For each epoch, we first compute the aggregated prediction score $\mathbf{H}_{b}$ of our boosted model as described in \cref{alg:inference}. We use a supervised exponential loss $\mathcal{L}_b$ for training with elastic-net regularization (L1 loss - $\mathcal L_1(.)$ and L2 loss - $\mathcal L_2(.) $) on the graph convolutional weight matrices $\mathbf{W}_{\mathbf{i}}^{k} \texttt{  }\forall\texttt{ } S_i \in \mathbf{S}$ for each view. Note that we employ weight sharing between all views of the same relation type so that only one set of weight matrices is learned per relation. 
The exponential loss, $\mathcal{L}_{\mathbf{R}}$, for each relation type is added alternatingly to the boosted loss.
We apply an \emph{exponential annealing schedule}, $\lambda(t)$, i.e. a function of the training epochs ($t$), to the loss function of each relation. As training progresses and the boosted model learns to optimally distribute vertices among the relations, increase in $\lambda(t)$ ensures more emphasis is provided to the individual convolutional networks of each relation. Figure \ref{fig:adaboost} illustrates the overall architecture of our IR-GCN model.

\section{Experiments}
\label{sec:experiments}
In this section, we first describe our dataset followed by our experimental setup; comparative baselines, evaluation metrics, and implementation details. We then present results across several experiments to evaluate the performance of our model on merging semantically diverse induced-relations. 
\begin{table*}[h]
 \centering
 \small
 \centering
 \begin{tabular}{l|c@{\hspace{0.8mm}}c@{\hspace{0.8mm}}c@{\hspace{0.8mm}}|c@{\hspace{1.2mm}}c@{\hspace{1.2mm}} c@{\hspace{1.2mm}} |c@{\hspace{1mm}}c@{\hspace{1mm}}c@{\hspace{1mm}}|c@{\hspace{1mm}} c@{\hspace{1mm}}c@{\hspace{1mm}}|c@{\hspace{1mm}}c@{\hspace{1mm}}c@{\hspace{1mm}} }
  \toprule
  &  \multicolumn{3}{c}{\textbf{Technology}} &
  \multicolumn{3}{c}{\textbf{Culture/Recreation}} &
  \multicolumn{3}{c}{\textbf{Life/Arts}} &
  \multicolumn{3}{c}{\textbf{Science}} &
  \multicolumn{3}{c}{\textbf{Professional/Business}}\\
  & ServerFault & AskUbuntu & Unix & English & Games & Travel & SciFi & Home & Academia & Physics & Maths & Statistics & Workplace & Aviation & Writing \\ \midrule
$\vert Q \vert$ & 61,873 & 41,192 & 9,207 & 30,616 & 12,946 & 6,782 & 14,974 & 8,022 & 6,442 & 23,932 & 18,464 & 13,773 & 8,118 & 4,663 & 2,932 \\
$\vert  \mathcal{A} \vert$ & 181,974 & 119,248 & 33,980 & 110,235 & 45,243 & 20,766 & 49,651& 23,956 & 23,837 & 65,800 & 53,772 & 36,022 & 33,220 & 14,137 & 12,009 \\
$ \vert U \vert$ & 140,676 & 200,208 & 84,026 & 74,592 & 14,038 & 23,304 & 33,754 & 30,698 & 19,088 & 52,505 & 28,181 & 54,581& 19,713 & 7,519 & 6,918 \\
$ \mu (\vert  \mathcal{A}_q \vert) $ & 2.94 & 2.89 & 3.69 & 3.6 & 3.49 & 3.06 & 3.31 & 2.99 & 3.7 & 2.75 & 2.91 & 2.62 & 4.09 & 3.03 & 4.10 \\
   \bottomrule
 \end{tabular}
 \caption{ \small \label{tab:stats}Dataset statistics for the top three Stack Exchange communities from five different categories. $\vert Q \vert$: number of questions; $\vert  \mathcal{A} \vert$: number of answers; $ \vert U \vert $: number of users; $ \mu (\vert  \mathcal{A}_q \vert) $: mean number of answers per question. Professional/Business communities have slightly more answers per question on average than others. Technology communities are the largest in terms of number of question out of the five categories.}
 \vspace{-0.2in}
\end{table*}

\subsection{Dataset}
We evaluate our approach on multiple communities catering to different topics from a popular online Community Question Answer (CQA) platform, \emph{StackExchange\footnote{https://stackexchange.com/}}. The platform divides the communities into five different categories, i.e. Technology ($\mathbf{T}$), Culture/Recreation ($\mathbf{C}$), Life/Arts ($\mathbf{L}$), Science ($\mathbf{S}$) and Professional ($\mathbf{P}$).
For our analysis, we collect data from the ten largest communities from each of the five categories until March 2019, resulting in a total of 50 StackExchange communities. In StackExchange, each questioner can mark a candidate answer as an "accepted" answer. We only consider questions with an accepted answer. Table \ref{tab:stats} shows the final dataset statistics.

For each $(q, a)$ tuple, we compute the following basic features:\\
\emph{Activity features :} View count of the question, number of comments for both question and answer, the difference between posting time of question and answer, arrival rank of answer (we assign rank 1 to the first posted answer) \cite{TianZL13}. \\
\emph{Text features :} Paragraph and word count of question and answer, presence of code snippet in question and answer (useful for programming based forums), word count in the question title.\\
\emph{User features :} Word count in user profile's Aboutme section for both users; one posting the question and other posting the answer.

Time-dependent features like upvotes/downvotes of the answer and user feature like reputation or badges used in earlier studies on StackExchange \cite{BurelMA16} are problematic for two reasons. First, we only know the aggregate values, not how these values change with time. Second, since these values typically increase over time, it is unclear if an accepted answer received the votes \emph{prior} to or \emph{after} an answer was accepted. Thus, we do not use such time-dependent features for our model and the baselines in our experiments.

\subsection{Experimental Setup}
\subsubsection{Baselines} We compare against state-of-the-art feature-based baselines for answer selection and competing aggregation approaches to fuse diverse relational views of the dataset~\cite{DualGCN, relationalGCN}.

\noindent
\textbf{Random Forest (RF)} \cite{BurelMA16, TianZL13} model trains on the feature set mentioned earlier for each dataset. This model is shown to be the most effective feature-based model for Answer Selection.

\noindent
\textbf{Feed-Forward network (FF)} \cite{JendersKN16} is used as a deep learning baseline to learn non-linear transformations of the feature vectors for each $(q, a)$ tuple. This model is equivalent to our Reflexive GCN model in isolation.

\noindent
\textbf{Dual GCN (DGCN)} \cite{DualGCN} trains a separate GCN for each view. In addition to the supervised loss computed using training labels, they introduce a regularizer to minimize mean squared error (MSE) between vertex representations of two views, thus aligning the learned latent spaces.
The regularizer loss is similar to our intra-relation aggregation approach but assumes label and feature sharing across \emph{all} the views.

\noindent
\textbf{Relational GCN (RGCN)} \cite{relationalGCN} combines the output representations of previous layer of each view to compute an aggregated input to the current layer.
%

We also report results for each view individually: Contrastive (C-GCN), Arrival Similarity (AS-GCN), TrueSkill Similarity (TS-GCN) and Reflexive (R-GCN) with our proposed IR-GCN model. We do not compare with other structure-based approaches to compute vertex representations \cite{DeepWalk, node2vec, Planetoid, LINE} as GCN is shown to outperform them \cite{gcn}. We also compare with common aggregation strategies to merge neural representations discussed earlier in ~\cref{sec:aggregation} later.

\subsubsection{Evaluation Metric}
We randomly select 20\% of the questions, $\mathbf{T}_q \subset \mathcal{Q}$ to be in the test set. Then, subsequently all $(q,a)$ tuples such that $q \in \mathbf{T}_q$ comprise the set of test tuples or vertices, $\mathbf{T}$ . The rest of the vertices, along with their label information, is used for training the model.
We evaluate our model on two metrics, Accuracy and Mean Reciprocal Rank (MRR). Accuracy metric is widely used in vertex classification literature while MRR is popular for ranking problems like answer selection. Formally,
\begin{align*}
Acc = \frac{1}{\vert \mathbf{T} \vert} \sum_{(q,a) \in  \mathbf{T} } \mathbbm{1} \left(  y_{(q,a)} \cdot h_b((q,a)) > 0 \right)
\end{align*}
with $\cdot$ as the product and $\mathbbm{1}$ as the indicator function. The product is positive if the accepted label and predicted label match and negative otherwise.
\begin{equation*}
MRR = \frac{1}{\vert \mathbf{T}_q \vert} \sum_{q \in \mathbf{T}_q} \frac{1}{\sum_{a' \in \mathcal{A}(q)}  \mathbbm{1} \left(L_{(q,a)} < L_{(q,a')} \right)} 
\end{equation*}
 where
$L_{(q,a)}$ refers to the position of accepted answer $a$ in the ranked list for question $q$ \cite{Wang:2009}.

 \begin{table*}[h]
   \robustify\bfseries
   \small
   \centering
   \begin{threeparttable}
  \begin{tabular}{l|S[round-mode=places,round-precision=2]@{\hspace{8mm}}S@{\hspace{10mm}}|S[round-mode=places,round-precision=2]@{\hspace{8mm}}S@{\hspace{10mm}}|S[round-mode=places,round-precision=2]@{\hspace{8mm}}S@{\hspace{10mm}}|S[round-mode=places,round-precision=2]@{\hspace{8mm}}S@{\hspace{10mm}}|S[round-mode=places,round-precision=2]@{\hspace{8mm}}S@{\hspace{10mm}}S[round-mode=places,round-precision=2]@{\hspace{8mm}}S@{\hspace{10mm}}}

     \toprule
     \multirow{2}{*}{Method} &
        \multicolumn{2}{c}{\textbf{Technology}} &
       \multicolumn{2}{c}{\textbf{Culture/Recreation}} &
       \multicolumn{2}{c}{\textbf{Life/Arts}} &
       \multicolumn{2}{c}{\textbf{Science}} &
       \multicolumn{2}{c}{\textbf{Professional/Business}}\\
       &{Acc(\%)} & {MRR}&{Acc(\%)} & {MRR}&{Acc(\%)}& {MRR}&{Acc(\%)} & {MRR}&{Acc(\%)} & {MRR}\\
       \midrule
     \textbf{RF~\cite{BurelMA16, TianZL13}} & 66.78 $\pm$ 0.023 & 0.6834 \ \ \   $\pm$ 0.043 & 72.5 $\pm$ 0.018 & 0.626\ \ \  $\pm$ 0.050 & 72.71 $\pm$ 0.049 & 0.6283\ \ \ $\pm$ 0.089 & 68.09 $\pm$ 0.024 & 0.6923\ \ \ $\pm$ 0.049 & 74.72 $\pm$ 0.044 & 0.5951\  \ \   $\pm$ 0.081\\

     \textbf{FF~\cite{JendersKN16}} & 67.31 $\pm$ 0.027 & 0.7860 \ \ \  $\pm$ 0.022 & 72.22134095 $\pm$ 0.020 & 0.782138231\ \ \  $\pm$ 0.023\textbf{*} & 73.57771352 $\pm$ 0.049 & 0.7801657493 \ \ \  $\pm$ 0.034 & 67.86881023 $\pm$0.024 & 0.7997881343 \ \ \  $\pm$ 0.028 & 74.63275816 $\pm$ 0.040 & 0.7595471154 \ \ \  $\pm$ 0.049\\
     \textbf{DGCN~\cite{DualGCN}} & 70.70 $\pm$ 0.022 & 0.7819\ \ \  $\pm$ 0.017 & 75.22 $\pm$ 0.017 & 0.7715263633\ \ \  $\pm$ 0.028 & 76.73 $\pm$ 0.034 &0.7840080442 \ \ \ $\pm$ 0.038 & 71.45 $\pm$ 0.023\textbf{*} & 0.7915379815\ \  \ $\pm$ 0.035 & 76.86 $\pm$ 0.031 & 0.7511047996\ \ \  $\pm$ 0.046\\
     \textbf{RGCN~\cite{relationalGCN}} & 54.40 $\pm$ 0.045 & 0.6726\ \ \ $\pm$ 0.045 & 60.39 $\pm$ 0.016 & 0.6455423723 \ \ \ $\pm$ 0.042 & 59.97 $\pm$ 0.043 & 0.654424422\ \ \  $\pm$ 0.054 & 58.65 $\pm$ 0.054 & 0.6825181603\ \ \  $\pm$ 0.042 &63.02 $\pm$ 0.038 & 0.6568131822\ \  \ $\pm$ 0.061\\
     \cmidrule(lr){1-1}\cmidrule(lr){2-11}
     \textbf{AS-GCN} & 67.76$\pm$ 0.032 &0.7746\ \  \ $\pm$0.015 & 73.05 $\pm$0.021 & 0.7630910769 \ \ \ $\pm$ 0.025 &73.79 $\pm$0.048 & 0.7765385098 \ \ \  $\pm$ 0.042 & 66.93 $\pm$ 0.045 & 0.7876480266 \ \ \ $\pm$0.028 & 74.99 $\pm$ 0.045 &0.7424021656 \ \ \ $\pm$0.047\\
     \textbf{TS-GCN} & 66.87 $\pm$ 0.032 & 0.7786 \ \ \ $\pm$ 0.018 & 72.16 $\pm$ 0.023 & 0.7642533243 \ \ \ $\pm$ 0.023 & 72.02 $\pm$ 0.061 & 0.7655344376\ \ \  $\pm$ 0.048 & 65.90 $\pm$ 0.042 & 0.7896596329\ \ \  $\pm$ 0.031 & 74.17 $\pm$ 0.046 &0.746963993 \ \ \ $\pm$ 0.044\\
     \textbf{C-GCN} & 71.64 $\pm$ 0.022\textbf{*} & 0.7902 \ \ \ $\pm$ 0.015 \textbf{*}& 76.18 $\pm$ 0.017\textbf{*}& 0.781 \ \ \ $\pm$ 0.024 & 77.37 $\pm$ 0.034 \textbf{*}& 0.7883908504 \ \ \ $\pm$ 0.040\textbf{*} & 70.81 $\pm$ 0.042 & 0.8002268874 \ \ \ $\pm$ 0.032 \textbf{*}& 77.57 $\pm$ 0.038\textbf{*} & 0.7679999323 \ \ \ $\pm$ 0.034\textbf{*}\\
     \textbf{IR-GCN} & \bfseries 73.96 $\pm$ \bfseries 0.023 & \bfseries0.7939\ \ \  $\pm$ \bfseries 0.014 & \bfseries 78.61 $\pm$ \bfseries 0.018 & \bfseries 0.7905241843 \ \ \ $\pm$ \bfseries 0.025 & \bfseries 79.21 $\pm$ \bfseries 0.032 & \bfseries 0.7995996677 \ \ \ $\pm$ \bfseries 0.037 & \bfseries 74.98 $\pm$ \bfseries 0.021 & \bfseries 0.8085288947 \ \ \ $\pm$ \bfseries0.028 & \bfseries80.17 $\pm$ \bfseries 0.026 & \bfseries 0.7849464934 \ \ \ $\pm$ \bfseries 0.032\\
     \bottomrule
   \end{tabular}
   \begin{tablenotes}
       \footnotesize
       \item[*] DGCN stands for DualGCN, RGCN stands for RelationalGCN, and IR-GCN stands for Induced Relational GCN.
   \end{tablenotes}
   \caption{\small \label{tab:stackacc} Accuracy and MRR values for StackExchange with state-of-the-art baselines. Our model outperforms by at least 4\% in Accuracy and 2.5\% in MRR. Contrastive GCN performs best among individual views. The model with $*$ symbol has the second-best performance among all other models. Our model shows statistical significance at level 0.01 overall second best model on single tail paired t-test.}
   \end{threeparttable}
 \vspace{-0.2in}
 \end{table*}

\subsubsection{Implementation Details}
We implemented our model and the baselines in Pytorch. We use ADAM optimizer \cite{ADAM} for training with 50\% dropout to avoid overfitting. We use four hidden layers in each GCN with hidden dimensions 50, 10, 10, 5, respectively, and ReLU activation. The coefficients of $\mathcal{L}_1$ and $\mathcal{L}_2$ regularizers are set to $\gamma_1 = 0.05$ and $\gamma_2 = 0.01$ respectively. For TrueSkill Similarity, we use margin $\delta = 4$ to create links, while for Arrival similarity, we use $\delta = 0.95$.
We implement a mini-batch version of training for large graphs where each batch contains a set of questions and their associated answers.  This is equivalent to training on the whole graph as we have disconnected cliques. All code and data will be released upon publication.

\subsection{Performance Analysis}
Table \ref{tab:stackacc} shows impressive gains over state-of-the-art baselines for all the five categories. We report mean results for each category obtained after 5-fold cross-validation on each of the communities. Our induced-relational GCN model beats best performing baseline by 4-5\% on average in accuracy. The improvement in MRR values is around 2.5-3\% across all categories. Note that MRR is based only on the rank of the accepted answer while accuracy is based on correct labeling of \emph{both} accepted and non-accepted answers.

\begin{figure}[h]
  \centering
    \vspace{-0.12in}
  \includegraphics[scale=0.35]{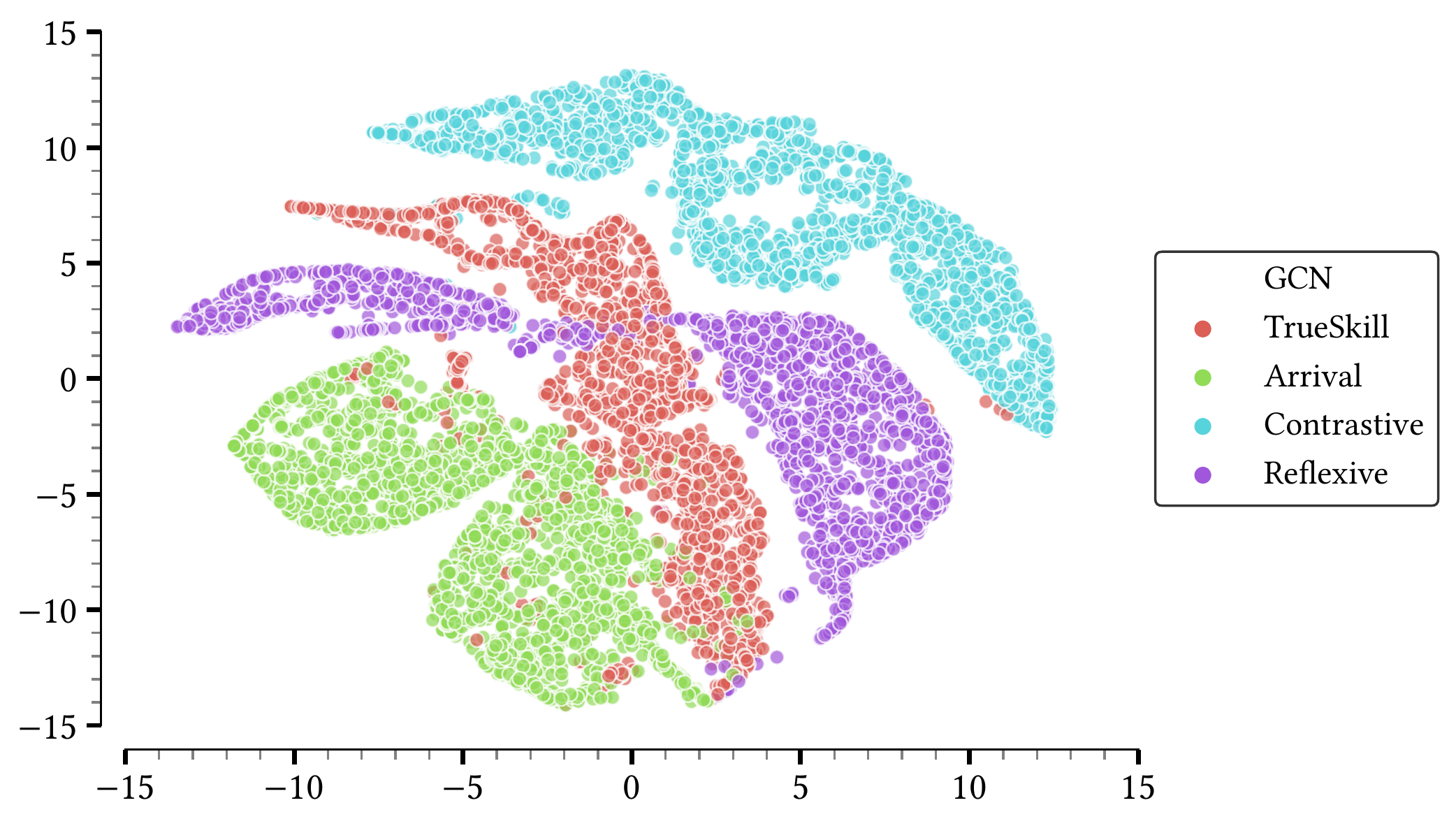}
  \caption{\small \label{fig:sne} t-stochastic neighbor embedding (t-SNE) \cite{sne} distributions of the learned vertex representations by our model for Chemistry StackExchange. Each view learns a distinct vertex representation. Best viewed in color.}
  \vspace{-0.1in}
\end{figure}

Among individual views, Contrastive GCN performs best on all the communities. It even beats the best performing baseline DualGCN that uses all the relational views. Note that contrastive view compares between the candidate answers to a question and uses our proposed contrastive modification to the convolution operation. Arrival Similarity follows Contrastive and then Reflexive. The superior performance of Arrival Similarity view shows that early answers tend to get accepted and vice versa. It indicates that users primarily use CQA forums for quick answers to their queries. Also, recall that Reflexive predicts each vertex's label independent of other answers to the same question. Thus, the competitive performance of Reflexive strategy indicates that vertex's features itself are well predictive of the label. TrueSkill Similarity performs at par or slightly worse than Reflexive. \Cref{fig:sne} presents t-SNE distributions \cite{sne} of the learned vertex representations ($\mathbf{Z}_i^K$) of our model applied to Chemistry StackExchange from Science category. Note that each view, including two views under Similar Contrast relation, learns a distinct vertex representation. Hence, all views are essential and contribute to our final performance.

Out of the baseline graph ensemble approaches, DualGCN performs significantly better than RelationalGCN by an average of around 26\% for all categories. Recall that in RelationalGCN model, the convolution output of each view is linearly combined to compute the final output. Linear combination works well for knowledge graphs as each view can be thought of as a feature, and then it accumulates information from each feature. DualGCN is similar to our approach and trains different GCN for each view and later merges their results. However, it enforces similarity in vertex representations learned by each view. This restriction is not suitable for our induced-relationships as they are semantically different (contrastive captures contrast in features vs. similarity enforces label sharing).

\subsection{Ablation Study on Relation Types}
\begin{table}[h]
  \vspace{-0.1in}
  \small
  \robustify\bfseries
  \begin{tabular}{l | c | c| c| c|c}
    \toprule
       \textbf{\{ Relation Type\}} &
        \textbf{Tech} &
        \textbf{Culture} &
        \textbf{Life} &
        \textbf{Sci}&
        \textbf{Business}\\
      \midrule
      C & 71.23 &75.90 &78.71&72.99 & 76.85\\
    \{ TS, AS \} & 67.86 &74.15 &75.75&65.80& 76.13  \\
    R & 68.30 & 73.35 & 76.57 & 67.40 & 75.76 \\
    \{TS, AS \} + R & 69.28 & 75.50 &76.41 &70.11  &77.90 \\
    C + R & 73.04 & 77.66 & 80.25 &73.72 & 80.04 \\
    C + \{ TS, AS \} & 72.81 & 78.04 & 81.41 & 72.19 & 80.15\\
    C + \{ TS, AS \} + R & \bfseries 73.87 & \bfseries 78.74 & \bfseries 81.60&  \bfseries74.68&  \bfseries80.56 \\
    \bottomrule
  \end{tabular}
  \caption{\small \label{tab:relation} 5-fold Accuracy (in \%) comparison for different combination of relation types for our boosted model. Contrastive and Similar Contrast relations together performs similar to the final model.}
  \vspace{-0.15in}
\end{table}

We present results of an ablation study with different combination of relation types (Contrastive, Similar and Reflexive) used for IR-GCN model in Table \ref{tab:relation}. We conducted this study on the biggest community from each of the five categories, i.e., ServerFault (Technology), English (Culture), Science Fiction (Life), Physics (Science), Workplace (Business).
Similar Contrast relation (TrueSkill and Arrival) used in isolation perform the worst among all the variants. Training Contrastive and Similar Contrast relation together in our boosted framework performs similar to our final model. Reflexive GCN contributes the least as it does not consider any neighbors.

\vspace{-0.1in}
\subsection{Aggregator Architecture Variants}
\label{sec:agg}
We compare our gradient boosting based aggregation approach with other popular methods used in literature to merge different neural networks discussed in \cref{item:aggregator}.
\begin{table}[h]
  \small
  \robustify\bfseries
  \begin{tabular}{l | c | c| c| c|c}
    \toprule
    \textbf{Method} &
     \textbf{Tech} &
     \textbf{Culture} &
     \textbf{Life} &
     \textbf{Sci}&
     \textbf{Business}\\
      \midrule
    Stacking~\cite{Stacking} &68.58 & 74.44 & 79.19 & 70.29 &75.50  \\
    Fusion~\cite{Fusion18}  &72.30 &77.25 & 80.79 & 73.91 &79.01 \\
    NeighborAgg~\cite{graphsage, relationalGCN}  &69.29 &74.28 & 77.94 & 68.42 &78.64   \\
    IR-GCN & \bfseries 73.87 & \bfseries 78.74 & \bfseries 81.60&  \bfseries74.78&  \bfseries80.56 \\
    \bottomrule
  \end{tabular}
  \caption{\small \label{tab:agg} 5-fold Accuracy (in \%) comparison of different aggregator architectures. These architectures perform worse than Contrastive GCN. Fusion performs similarly but is computationally expensive.}
  \vspace{-0.2in}
\end{table}

Table \ref{tab:agg} reports the accuracy results for these aggregator variants as compared to our model. Our method outperforms all the variants with Fusion performing the best.  This reaffirms that existing aggregation models are not suitable for our problem. Note that these approaches perform worse than even Contrastive GCN except Fusion. Fusion approach performs similar to our approach but is computationally expensive as the input size for each IR-GCN is linear in the number of all views in the model.

\subsection{Discriminative Magnification effect}

\begin{figure}[b]
  \centering
  \includegraphics[scale=0.3]{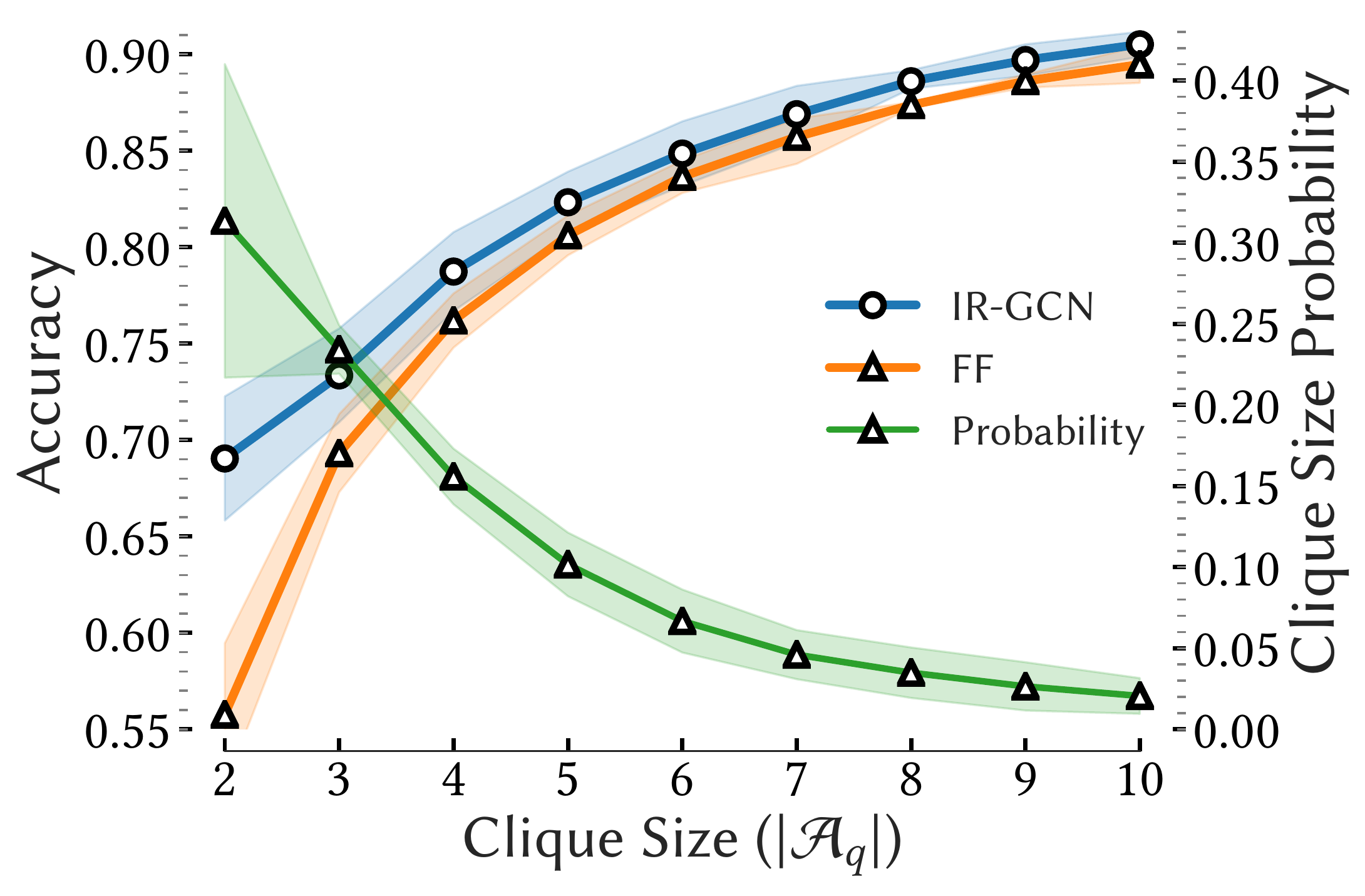}
  \vspace{-0.12in}
  \caption{\small \label{fig:clique} Accuracy of our IR-GCN model compared to the FF model with varying clique size (i.e. number of answers to a question, $\vert \mathcal{A}_q \vert$) for Contrastive view . 
We report averaged results over the largest community of all categories. Our model performs much better for smaller cliques, and the effect diminishes for larger cliques (\cref{eq:contrast}). 80\% of the questions have $< 4$ answers.}
  \vspace{-0.15in}
\end{figure}

We show that due to our proposed modification to the convolution operation for contrastive view, we achieve \emph{Discriminative Magnification effect} (\cref{eq:contrast}). Note that the difference is scaled by Clique size ($1 + 1/n-1$), i.e. number of answers to a question, $\vert \mathcal{A}_q \vert$. Figure \ref{fig:clique} shows the accuracy of our IR-GCN model as compared to the FeedForward model with varying clique size. Recall that FeedForward model predict node labels independent of other nodes and is not affected by clique size. We report average results over the same five communities as above. We can observe that increase in accuracy is much more for lower clique sizes (13\% improvement for $\vert \mathcal{A}_q \vert = 2$ and 4\% for $\vert \mathcal{A}_q \vert = 3$ on average). The results are almost similar for larger clique sizes. In other words, our model significantly outperforms the FeedForward model for questions with fewer candidate answers. However, around 80\% of the questions have very few answers($< 4$) and thus this gain over FF is significant.

\subsection{Label Sparsity}

\begin{figure}
      \centering
  \includegraphics[scale=0.3]{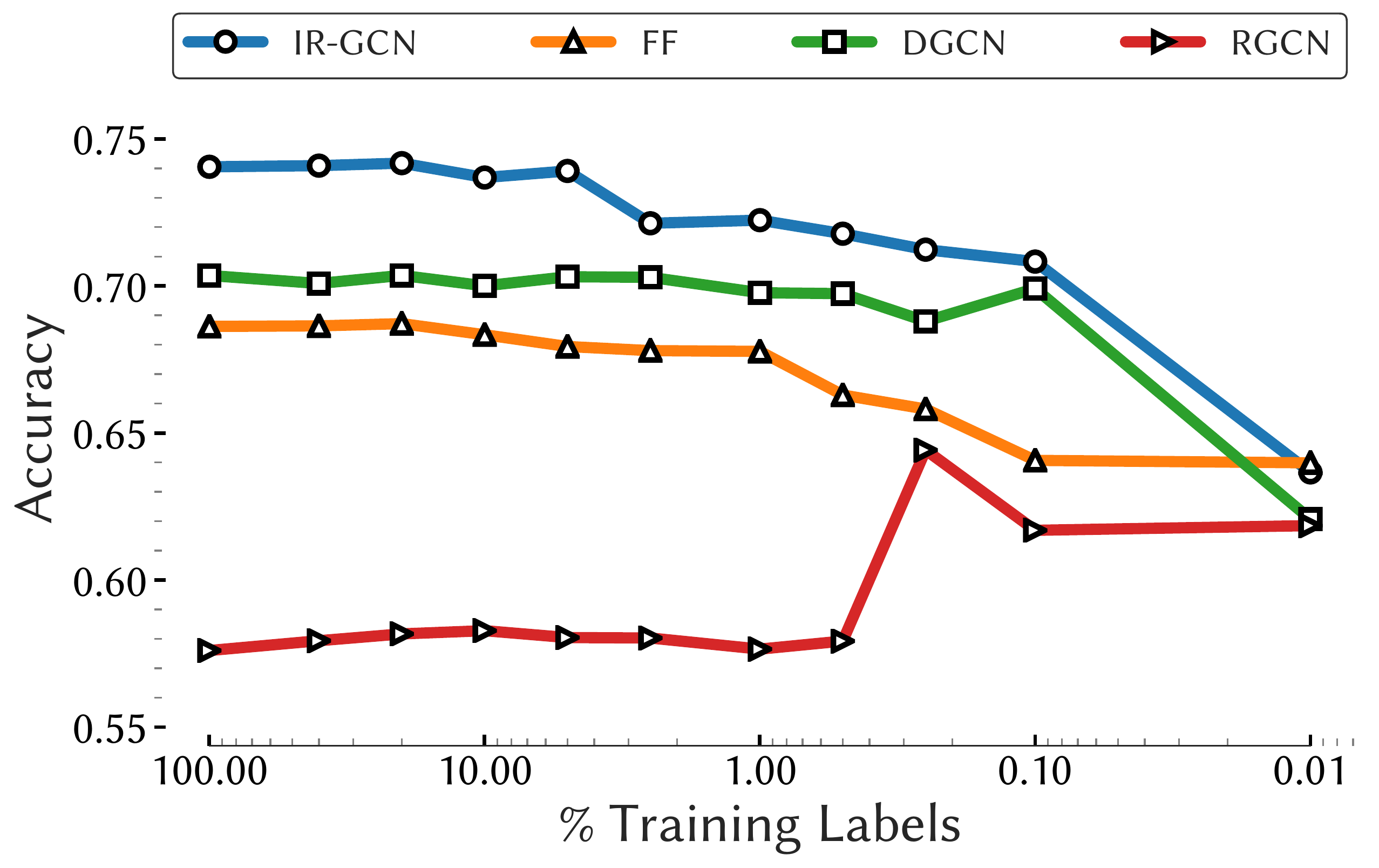}
  \vspace{-0.2in}
  \caption{\small\label{fig:labelsparsity} Change in accuracy with varying training label rates for Physics StackExchange. Our model is more robust to label sparsity than other relation ensemble approaches. RGCN works better with fewer labels as contrastive relation introduces noise in the model. At extreme sparsity, all approaches converge to the same value indicating random selection.}
  \vspace{-0.2in}
\end{figure}
Graph Convolution Networks are robust to label sparsity as they exploit graph structure and are thus heavily used for semi-supervised settings. Figure \ref{fig:labelsparsity} shows the change in accuracy for Physics StackExchange from Science category at different training label rates. Even though our graph contains disconnected cliques, IR-GCN still preserves robustness to label sparsity.
In contrast, the accuracy of FeedForward model declines sharply with less label information. Performance of DualGCN remains relatively stable while Relational GCN's performance increases with a decrease in label rate. Relational GCN assumes each view to be of similarity relation and thus, adding contrastive relation introduces noise in the model. However, as the training labels become extremely sparse, the training noise decreases that leads to a marked improvement in the model. In case of extremely low label rate of 0.01\%, all approaches converge to the same value, which is the expectation of theoretically random selection. We obtained similar results for other four StackExchange communities but omitted them for brevity.

\subsection{Limitations} We do recognize certain limitations of our work. First, we do not deal with content in our model. Our focus in this work is to exploit structural properties between tuples. We believe that adding content will further improve our results. Second, we focus on equivalence relations that induce a graph comprising cliques. While cliques are useful graph objects for answer selection, equivalence relations may be too restrictive for other problems (e.g., the relation is not transitive). However, our modular framework does apply to arbitrary graphs, except that~\Cref{eq:restrictk} will no longer be an \emph{exact} convolution but be an approximation. 

\section{Related Work}
\label{sec:related}
Our work intersects two research areas; Answer Selection and handling multi-relational social data, primarily via Graph Convolution.

\textbf{Answer Selection}
In CQA forums, previous answer selection literature includes feature-driven models and deep text models.

\noindent
\emph{Feature-Driven Models \cite{BurelMA16,  JendersKN16, TianZL13, TianL16}} in CQA identify and incorporate user features, content features, and thread features, e.g., in tree-based models \cite{BurelMA16, JendersKN16, TianZL13} to identify the best answer. \citet{TianZL13} found that the best answer tend to be early and novel, with more details and comments. \citet{JendersKN16} trained classifiers for online forums, \citet{BurelMA16} emphasize the Q\&A thread structure.  


\noindent
\emph{Deep Text Models \cite{ZhangLSW17, WuWS18, WangN15, SukhbaatarSWF15}} learn optimal QA text-pair representations to select the best answer. \citet{FengXGWZ15} augment CNNs with discontinuous convolution for improved representations; \citet{WangN15} use stacked biLSTMs to match question-answer semantics.

\textbf{Multi-Relational Approaches:} While text and feature models treat answer content independently, we focus on integrating multi-relational social aspects in the prediction. We identify a few related threads; adversarial approaches to integrate social neighbor data~\cite{adv_social, adv_neighbor}; meta-learning to adapt across data modalities or tasks~\cite{maml}. Different from these directions, we focus on the simplicity of our multi-relational graph formulation.


\textbf{Graph Convolution} is applied in spatial and spectral domains to compute graph node representations for downstream tasks including node classification \cite{gcn}, link prediction \cite{relationalGCN}, multi-relational tasks~\cite{rase} etc. Spatial approaches employ random walks or k-hop neighborhoods to compute node representations \cite{DeepWalk, node2vec, Planetoid, LINE} while fast localized convolutions are applied in the spectral domain\cite{deferrard, duvenaund}. Our work is inspired by Graph Convolution Networks (GCN) \cite{gcn}, which outperforms spatial convolutions and scales to large graphs. GCN extensions have been proposed for signed networks \cite{signedgcn}, inductive settings \cite{graphsage}, multiple relations \cite{DualGCN, relationalGCN} and evolution~\cite{dysat}. However, GCN variants assume label sharing, which cannot model contrastive relations in our setting. 


\section{Conclusion}
\label{sec:conclusion}

This paper addressed the question of identifying the accepted answer to a question in CQA forums. We developed a novel induced relational graph convolutional (IR-GCN) framework to address this question. We made three contributions. First, we introduced a novel idea of using strategies to induce different views on $(q,a)$ tuples in CQA forums. Each view consists of cliques and encodes---reflexive, similar, contrastive---relation types. Second, we encoded label sharing and label contrast mechanisms within each clique through a GCN architecture.  Our novel contrastive architecture achieves \emph{Discriminative Magnification} between nodes. Finally, we show through extensive empirical results on StackExchange that boosting techniques improved learning in our convolutional model.
Our ablation studies show that the contrastive relation is most effective individually in StackExchange. As part of future work, we plan to include content into the convolutional framework.


\printbibliography
\end{document}